\documentstyle[preprint,eqsecnum,aps]{revtex}

\begin{document}
\title{Autonomous generation of all Wigner functions and marginal probability
densities of Landau levels}
\author{B. Demircio\u glu$^{1}$ \thanks{%
E.mail:bengu.demircioglu@taek.gov.tr} and A. Ver\c{c}in$^{2}$\thanks{%
E.mail:vercin@science.ankara.edu.tr}}
\address{$^{1}$ Ankara Nuclear Research and Training Center\\
06100, Be\c{s}evler, Ankara-Turkey\\
$^{2}$ Department of Physics, Ankara University, Faculty of Sciences,\\
06100, Tando\u gan-Ankara, Turkey\\
}
\maketitle

\begin{abstract}
Generation of Wigner functions of Landau levels and determination of their
symmetries and generic properties are achieved in the autonomous framework
of deformation quantization. Transformation properties of diagonal Wigner
functions under space inversion, time reversal and parity transformations
are specified and their invariance under a four-parameter subgroup of
symplectic transformations are established. A 
generating function for all Wigner functions is developed and this has been
identified as the phase-space coherent state for Landau levels. Integrated
forms of generating function are used in generating explicit expressions of
marginal probability densities on all possible two dimensional phase-space
planes. Phase-space realization of unitary similarity and gauge
transformations as well as some general implications for the Wigner function
theory are presented.
\end{abstract}

\tightenlines

\section{Introduction}

Landau levels are equally spaced, highly degenerate energy levels of a
charged particle moving on a plane under the influence of a perpendicular
uniform and static magnetic field \cite{Landau,Aoki}. They play very
important roles in our current understanding of two dimensional electron
systems and their characteristic behaviors such as quantum Hall effect \cite
{Aoki,Isihara,Laughlin} and superconductivity \cite{Tesanovic} and in
exploring electronic properties of various bound states of matter such as
atoms, molecules and condensed matter in strong magnetic fields \cite{Lai}.
Regarding recent experiments \cite{Nature,Leibfried} devised to measure
Wigner function of various states of light and matter and to observe its
remarkable properties, determination of Wigner functions of Landau levels
and search for their salient features become very promising since they can
be realized in several classes of systems.

Wigner function is quantum mechanical analogue of classical phase-space
probability distribution function. It is the central concept of
Weyl-Wigner-Groenewold-Moyal (WWGM) quantization which was initiated just
after the formulation of quantum mechanics \cite{Hillery}. WWGM quantization
consists of well established correspondence rules between the operator
formulation of quantum mechanics and its phase-space formulation \cite
{Hillery,Cahill,Vercin1}. According to these rules Wigner function can be
computed from 
\begin{equation}
W_{\lambda \lambda ^{\prime }}=\int_{{\bf R}^{D}}\psi
_{\lambda }({\bf q}+\frac{1}{2}{\bf y})\overline{\psi}_{\lambda ^{\prime }}({\bf q%
}-\frac{1}{2}{\bf y})e^{-i\frac{{\bf y}\cdot {\bf p}}{\hbar }}dV(y),
\end{equation}
where, $\hat{H}$ being the Hamiltonian operator, $\psi _{\lambda }$ is an
eigenfunction of the Schr\"{o}dinger equation $\hat{H}\psi_{\lambda}=E_{%
\lambda }\psi_{\lambda }$, $\overline{\psi}$ denotes the complex conjugation of $%
\psi$ and $({\bf q},{\bf p})\equiv (q_{1},..., q_{D},p_{1},..., p_{D})$ are
the phase space coordinates. $dV(y)$ stands for $D$-dimensional volume
element in $y_{j}$ variables, $\hbar =h/2\pi $ is the Planck's constant, and 
${\bf y}\cdot {\bf p}$ denotes the usual scalar product of vectors ${\bf y}$
and ${\bf p}$. The functions
$W_{\lambda \lambda ^{\prime }}\equiv
W_{\lambda \lambda ^{\prime }}({\bf q},{\bf p})$ are called
off-diagonal Wigner functions and what is referred to as Wigner function is
known also as the diagonal (or, pure state) Wigner function which
corresponds taking $\psi_{\lambda }=\psi _{\lambda ^{\prime }}$ in (1.1),
where $\lambda $ may be a multi index set of quantum numbers. Throughout
this paper the pair $(q_{j},p_{j})$ will denote the canonically conjugate
position and momentum variables for each $j=1,2,...,D$, bold face letters
will stand for vector quantities and when there is no risk of confusion
arguments of functions will be suppressed.

Deformation quantization which was formulated in the pioneering
work \cite{Bayen} in the seventies by generalizing WWGM quantization
has become the third autonomous and logically complete formulation
of quantum mechanics beyond the conventional ones based on operators
in Hilbert space or path integral \cite{Sternheimer,Lichnerowicz}. In
this formulation quantum effects are encoded in a new operation
called $\star $-product. This is a noncommutative but associative
composition rule between functions defined on phase-space which
can be any symplectic manifold \cite{Lichnerowicz,Marsden} (see
also \cite{Sternheimer,Kontsevich} for
recent developments) . In terms of $\star $-product the spectrum
and corresponding phase-space eigenfunctions of a Hamiltonian
function $H$ can be determined through
two-sided $\star $-eigenvalue equation \cite{Fairly1,Fairly,Curtright} 
\begin{equation}
H\star W_{\lambda }=W_{\lambda }\star H=E_{\lambda }W_{\lambda }.
\end{equation}
In that case the Wigner function of $H$ corresponding to the eigenvalue $%
E_{\lambda }$ is the $\star $-eigenfunction $W_{\lambda }$ of (1.1). When
the underlying phase-space is ${\bf R}^{2D}$ with globally
defined coordinates $({\bf q},{\bf p})$, the $\star $-product is given by 
\begin{equation}
\star =\exp [\frac{1}{2}i\hbar \sum_{j=1}^{D}(\stackrel{\leftarrow }{%
\partial }_{q_{j}}\stackrel{\rightarrow }{\partial }_{p_{j}}-\stackrel{%
\leftarrow }{\partial }_{p_{j}}\stackrel{\rightarrow }{\partial }_{q_{j}})],
\end{equation}
where we use the abbreviation $\partial _{x_{i}}\equiv \partial /\partial
x_{i}$ and the convention that $\stackrel{\leftarrow }{\partial }$ and $%
\stackrel{\rightarrow }{\partial }$ are acting, respectively, on the left
and on the right. (1.3) is known as the Moyal $\star$-product.

Apart from a few simple cases the use of (1.1) to obtain Wigner
function is rather difficult since for a given wavefunction the
resulting integrals are not easy to cope with. On the other
hand, only a few $\star $-eigenvalue problems have been analytically
handled until now. In fact, the basic physical systems such as H atom
and harmonic oscillator were considered in \cite{Bayen} with special
emphasis on determination of their spectra. Later on, some
studies utilizing different techniques considered the same
and some other systems in phase-space formulation
\cite{Fairly1,Fairly,Curtright,Gracia,Man'ko}. To the best of our
knowledge, the Ref.\cite{Man'ko} is the first in dealing with
Wigner function of Landau levels by using some WWGM correspondence
rules. Systematic use of $\star $-eigenvalue equations were
taken up in Ref.\cite{Curtright} which inspired us much.

The main goal of the present paper is to generate all Wigner functions and
corresponding marginal probability densities of Landau levels on all
possible phase-space planes. In doing that we shall follow an autonomous
approach by working entirely within the framework of deformation
quantization. All generic properties such as reality, normalization,
projection and orthogonality as well as transformation and symmetry
properties of these bound state Wigner functions are rigorously investigated in this
framework without any reference to wavefunctions or to WWGM correspondence
rules. After solving a $\star$-eigenvalue equation for the ground state
Wigner function, Wigner functions of the excited states will be produced
algebraically by successive applications of creation and annihilation
functions. Then, off-diagonal Wigner functions and marginal probability
densities will be obtained by means of associated generating functions.

The organization of the paper is as follows. In the next section the main
points of deformation quantization, mainly those needed in subsequent
investigation are briefly reviewed. Formulation of the problem and a closed
algebraic form of diagonal Wigner functions of Landau levels are given in
Sec. III, and their explicit functional forms are derived in Secs. IV and V.
Symmetries and other generic properties are taken up in Secs. VI and VII. A
generating function is introduced in Sec. VIII and all (diagonal and
off-diagonal) Wigner functions are generated by means of it. In Sec. IX
phase-space coherent state property of this generating function is presented
and it is shown that its integrated forms serve as generating functions for
marginal probability densities derived in Sec. X and presented altogether in
Tables I and II. Final section is devoted to some remarks concerning
justifications of some our results, implementation of unitary similarity
transformations and gauge transformations in a phase space and to some
general implications for Wigner function theory. In appendix we derive an
operator identity used in the main text and exhibit a new finite sum formula
for the generalized Laguerre polynomials.

\section{Star Product and Moyal Bracket}

Let us denote the linear space of complex valued, smooth (differentiable to
all orders) functions defined on ${\bf R}^{2D}$ by ${\cal N} $. The Moyal $%
\star $-product (1.3) is a bilinear product $\star : {\cal N}\times {\cal N}%
\rightarrow {\cal N} $, which is, for all $f_{j} \in {\cal N}$, associative 
\begin{equation}
f_{1}\star (f_{2}\star f_{3})=(f_{1}\star f_{2})\star f_{3},
\end{equation}
and obeys the so-called closedness property \cite{Fairly1,Connes} 
\begin{equation}
\int_{{\bf R}^{2D}} f_{1} \star f_{2}dV = \int_{{\bf R}^{2D}} f_{1}f_{2}dV
=\int_{{\bf R}^{2D}} f_{2} \star f_{1}dV.
\end{equation}
This can be proved via integration by parts and it implies that integration,
all over the phase-space, is a trace on the $\star$-algebra of functions.
From (1.3) immediately follows that under complex conjugation the $\star$%
-product of functions satisfies the following relation 
\begin{equation}
\overline{(f_{1}\star f_{2})}=\overline{f}_{2}\star \overline{f}_{1}.
\end{equation}
It is also evident that $f_{1}({\bf q})\star f_{2}({\bf q})=f_{1}({\bf q})
f_{2}({\bf q})$ and $g_{1}({\bf p})\star g_{2}({\bf p})=g_{1}({\bf p}) g_{2}(%
{\bf p})$, where the product on the right hand sides is the usual
commutative pointwise product of functions.

In terms of $\star $-product Moyal bracket $\{,\}_{M}$ is
defined, for all $f,g\in {\cal N}$, as 
\begin{equation}
\{f,g\}_{M}=f\star g-g\star f.
\end{equation}
The properties of the $\star $-product ensure that Moyal bracket (MB) is
bilinear, antisymmetric and obey the Jacobi identity and Leibniz rule. With
respect to MB, ${\cal N}$ acquires a Lie algebra structure. The most
important properties of $\star $-product and MB are the following limit
relations 
\begin{mathletters}
\begin{eqnarray}
\lim_{\hbar \rightarrow 0}f\star g&=&fg,\\
\lim_{\hbar \rightarrow
0}(i\hbar )^{-1}\{f,g\}_{M}&=&\{f,g\}_{P},
\end{eqnarray}
\end{mathletters}
where
\begin{equation}
\{f,g\}_{P}=\sum_{k=1}^{D}(\partial _{q_{k}}f\partial _{p_{k}}g-\partial
_{p_{k}}f\partial _{q_{k}}g),
\end{equation}
is the usual Poisson bracket (PB) of the classical mechanics. The limit
relations (2.5) reveal the fact that the associative $\star $-algebra and
Lie algebra structure of ${\cal N}$ given by MB are, respectively,
deformations (in the sense of Gerstanhaber\cite{Gerstenhaber}) of
associative algebra structure of ${\cal N}$ with respect to above mentioned
pointwise product and of Lie algebra structure determined with respect to PB.

The set ${\cal A}$ of phase-space functions $f$'s which satisfy 
\begin{equation}
\{f,g\}_{M}=i\hbar \{f,g\}_{P},
\end{equation}
for all $g\in {\cal N}$ plays an important role in deformation quantization.
First, ${\cal A}$ is a Lie subalgebra with respect to PB which is preserved
under the deformation $PB\rightarrow MB$. Second, with respect to PB, ${\cal %
A}$ acts as a derivation on the $\star $-algebra of functions, that is, 
\begin{equation}
\{f,g_{1}\star g_{2}\}_{P}=\{f,g_{1}\}_{P}\star g_{2}+g_{1}\star
\{f,g_{2}\}_{P},
\end{equation}
for all $g_{1},g_{2}\in {\cal N}$ and $f\in {\cal A}$. Finally, quantum
mechanical time evolution of any element of ${\cal A}$ is a classical
evolution. This implies that when the Hamiltonian function belongs to ${\cal %
A}$, the classical and quantum time evolutions of an observable coincide. In
the nomenclature of the deformation quantization ${\cal A}$ is called the
invariance algebra of the $\star $-product and its elements are called
preferred (good or distinguished) observables. As can easily be verified
from (1.3) and (2.4) ${\cal A}$ is spanned by 
\begin{equation}
\{1,\ q_{j},\ p_{j},\ q_{j}q_{k\geq j},\ p_{j}p_{k\geq j},\ q_{j}p_{k}\}.
\end{equation}
These define $2D^{2}+3D+1$ dimensional affine symplectic algebra $%
w_{D}\oplus sp(2D)$, where $w_{D}$ stands for $2D+1$ dimensional
Heisenberg-Weyl algebra and $sp(2D)$ denotes $2D^{2}+D$ dimensional
symplectic algebra.

For $g={\bf c}_{1}\cdot {\bf p}+f$, where ${\bf c}_{1}$ is a constant vector
and $f=f({\bf q})$, we have 
\begin{eqnarray}
g\star g^{n}=g^{n+1}+ \sum_{j=2}^{\infty} \frac {1}{j!}(\frac {i\hbar}{2}%
)^{j} f( \sum_{k=1}^{D} \stackrel{\leftarrow}{\partial}_{q_{k}} \stackrel{%
\rightarrow}{\partial}_{p_{k}})^{j} g^{n}.  \nonumber
\end{eqnarray}
Hence, if $f$ is first order in $q_{k}$'s then $g\star g^{n}=g^{n+1}$ for
any $n$, which implies that 
\begin{eqnarray}
(g_{\star})^{n}=g^{n},\quad {\rm for} \quad g =c_{0}+ {\bf c}_{1}\cdot {\bf %
p}+{\bf c}_{2}\cdot {\bf q},
\end{eqnarray}
where $c_{0}$ is a constant, ${\bf c}_{2}$ is another constant vector and we
have defined the $\star $-power of $g\in {\cal N}$ as 
\begin{eqnarray}
(g_{\star})^{n}\equiv g\star g\star \cdots \star g , (n \ {\rm times}).
\end{eqnarray}
Finally in this section we should note that for two functions of the form $%
f=a_{1}+h_{1}, g=a_{2}+h_{2}$ such that $a_{1}, a_{2}\in {\cal A} $, and $%
\{h_{1}, h_{2}\}_{M}= i\hbar\{h_{1}, h_{2}\}_{P} $ we have $\{f,
g\}_{M}=i\hbar \{f, g\}_{P}$. As particular cases this holds for $%
h_{j}=h_{j}({\bf p})$, or for $h_{j}=h_{j}({\bf q})$.

\section{Wigner Functions of Landau Levels}

The well known Landau Hamiltonian $H_{L}$ for a spinless particle of charge $%
q>0$ and mass $m$ moving on the $q_{1}q_{2}$-plane is (in the Gaussian
units) 
\begin{equation}
H_{L}=\frac{1}{2m}({\bf p}-\frac{q}{c}{\bf A})^{2}=\frac{1}{2}%
m(v_{1}^{2}+v_{2}^{2}),
\end{equation}
where $c$ is the speed of light, ${\bf A}\equiv {\bf A}({\bf q})$ is the
vector potential of the magnetic field 
\begin{equation}
B=\partial _{q_{1}}A_{2}-\partial _{q_{2}}A_{1},
\end{equation}
and ${\bf v}=({\bf p}-\frac{q}{c}{\bf A})/m$ is the velocity vector whose
components obey 
\begin{equation}
\{v_{1},v_{2}\}_{M}=i\frac{\hbar q}{m^{2}c}B.
\end{equation}
In writing (3.3) we have used the fact that the phase-space of the Landau
problem is ${\bf R}^{4}$ with canonical coordinates $({\bf q},{\bf p})$. In
that case $\star $-product and MB are given by (1.3) and (2.4) for $D=2$. We
should also note that (3.3) is valid for any $B$.

When the magnetic field is constant the general solution of (3.2) is 
\begin{equation}
{\bf A}=(-\frac{1}{2}Bq_{2}+\partial _{q_{1}}\chi ,\frac{1}{2}%
Bq_{1}+\partial _{q_{2}}\chi),
\end{equation}
where $\chi \equiv \chi ({\bf q})$ is an arbitrary gauge function. In such a
case we have, in any gauge $\chi $, two constants of motion 
\begin{mathletters}
\begin{eqnarray}
X_{1}&=&m(v_{2}+\omega q_{1}),\\
X_{2}&=&-m(v_{1}-\omega q_{2}),
\end{eqnarray}
\end{mathletters}
where $\omega =qB/mc$ is the cyclotron frequency. ($X_{1}/m\omega
,X_{2}/m\omega $) correspond to the coordinates of cyclotron centre and
satisfy the following gauge-independent relations 
\begin{mathletters}
\begin{eqnarray}
\{X_{1},X_{2}\}_{M} &=&-im\hbar \omega , \\
\{v_{j},X_{k}\}_{M} &=&0=\{H_{L},X_{k}\}_{M},\quad j,k=1,2.
\end{eqnarray}
The relations $\{H_{L},X_{k}\}_{M}=0$ imply that $H_{L}, X_{1}$
and $X_{2}$ are constants of motion and since 
\end{mathletters}
\begin{equation}
dX_{1}\wedge dX_{2}\wedge dH_{L}=m^{3}\omega \lbrack
(v_{1}dq_{2}-v_{2}dq_{1})\wedge dv_{1}\wedge dv_{2}+\frac{\omega }{m}{\bf v}%
\cdot {\bf dp}\wedge dq_{1}\wedge dq_{2}]
\end{equation}
they are functionally independent provided that ${\bf v}\neq 0$. In Eq.
(3.7) $dq_{j},dp_{j}$ denote the coordinate differentials (coordinate
1-forms) and $\wedge $ stands for the antisymmetric exterior product \cite
{Marsden}. The above observations emphasize the quantum superintegrability
of the problem. Moreover, by the remark at the end of previous section, it
is easy to check that MBs in Eqs. (3.3) and (3.6) are equal to $i\hbar $
times the corresponding Poisson brackets. This implies that $%
\{H_{L},X_{1},X_{2}\}$ are at the same time classical constants of motion in
any gauge, and hence the problem is also classically superintegrable. As a
result, the Landau problem is one of the rare problems for which the
classical and quantum superintegrability coincide.

For the phase-space quantization of the Landau problem we shall use $H_{L}$
and 
\begin{equation}
J=\frac{1}{2m\omega }(X^{2}-m^{2}v^{2}),
\end{equation}
as a complete Moyal-commuting phase-space functions. $J$ corresponds to
canonical angular momentum $q_{1}p_{2}-q_{2}p_{1}$ in the symmetric gauge $%
\chi=$constant. We then introduce two mutually commuting pairs of
dimensionless creation and annihilation functions 
\begin{mathletters}
\begin{eqnarray}
a &=&\frac{1}{\gamma \omega }(v_{1}+iv_{2}),\nonumber\\
\bar{a}&=&\frac{1}{\gamma
\omega }(v_{1}-iv_{2}), \\
b &=&\frac{1}{m\gamma \omega }(X_{2}+iX_{1}),\nonumber\\
\bar{b}&=&\frac{1}{m\gamma
\omega }(X_{2}-iX_{1}),
\end{eqnarray}
where $\gamma =(2\hbar c/qB)^{1/2}=(2\hbar /m\omega)^{1/2}$ is the so-called
magnetic length. $\gamma $ corresponds the radius of a disc from which a
flux quantum $hc/e$ passes for $q=e$, $e$ being the elementary charge. The
nonvanishing Moyal brackets of (3.9) are 
\end{mathletters}
\begin{eqnarray}
\{a,\bar{a}\}_{M} =1=\{b,\bar{b}\}_{M}.
\end{eqnarray}
In terms of real number functions defined by 
\begin{eqnarray}
N_{a}=\bar{a}\star a,\quad N_{b}=\bar{b}\star b,
\end{eqnarray}
(3.1) and (3.8) take the form 
\begin{mathletters}
\begin{eqnarray}
H_{L} &=& \hbar \omega (N_{a}+\frac{1}{2}),\\
J &=& \hbar (N_{b}-N_{a}).
\end{eqnarray}
\end{mathletters}

By making use of
\begin{mathletters}
\begin{eqnarray}
\{a, (\bar{a}_{\star})^{k}\}_{M} &=&k(\bar{a}_{\star})^{k-1},\\
 \{b, (\bar{b}_{\star})^{k} \}_{M}&=&k(\bar{b}_{\star})^{k-1} ,
\end{eqnarray}
\end{mathletters}
and by defining the ground state Wigner function $W_{0}$ by
\begin{eqnarray}
a\star W_{0} = 0 = b\star W_{0},
\end{eqnarray}
it is straightforward to check that 
\begin{eqnarray}
W_{nl} = \frac{1}{n ! l !} (\bar{a}_{\star})^{n}\star (\bar{b}%
_{\star})^{l}\star W_{0} \star (a_{\star})^{n}\star (b_{\star})^{l},
\end{eqnarray}
satisfy the following ``$\star$-ladder structure" 
\begin{mathletters}
\begin{eqnarray}
a\star W_{nl} &=& W_{n-1, l}\star a,\nonumber\\
b\star W_{nl} &=& W_{n,l-1}\star b,\\
\bar{a}\star W_{nl} &=& W_{n+1, l}\star \bar{a},\nonumber\\
\bar{b}\star W_{nl}&=&W_{n, l+1}\star \bar{b},
\end{eqnarray}
where $n,l$ are positive integers. We assume that $W_{0}\equiv W_{00}$
is a real function. Eqs.
(3.16b) can also be verified by taking the complex conjugates of (3.16a) and
then performing the shifts $n\rightarrow n+1$ and $l \rightarrow l+1$. For
the number functions (3.11) we have 
\end{mathletters}
\begin{mathletters}
\begin{eqnarray}
N_{a}\star W_{nl} &=& W_{nl}\star N_{a} = \bar{a}\star W_{n-1, l}\star a =
nW_{nl}, \\
N_{b}\star W_{nl} &=& W_{nl}\star N_{b} = \bar{b}\star W_{n, l-1}\star b = l
W_{nl}.
\end{eqnarray}
Hence, $\{ W_{nl} : n,l=0, 1,... \}$
is the set of simultaneous eigenfunctions of
$N_{a}$ and  $N_{b}$ whose spectra are bounded from below.

In view of the above relations it is now easy to show that $W_{nl}$ satisfy 
\end{mathletters}
\begin{mathletters}
\begin{eqnarray}
H_{L}\star W_{nl} &=&W_{nl}\star H_{L}=E_{n}W_{nl}, \\
J\star W_{nl} &=&W_{nl}\star J=J_{nl}W_{nl},
\end{eqnarray}
with $\star $-eigenvalues 
\end{mathletters}
\begin{mathletters}
\begin{eqnarray}
E_{n} &=&\hbar \omega (n+\frac{1}{2}), \\
J_{nl} &=&\hbar (l-n),\quad n,l=0,1,2,...
\end{eqnarray}
$E_{n}$'s are the well-known infinitely degenerate ( since every $E_{n}$ is
independent from the quantum number $l$) Landau levels. The corresponding
Wigner functions are given by (3.15) (or, by Eq. (3.22) below) in a closed
algebraic way and in a gauge-independent manner. We should note that the
degeneracy of a Landau level is a reflection of the fact that energy is
independent of the location of the guiding center of gyrating charge. It is
infinite when the motion takes place on the whole ${\bf R}^{2}$ plane and
otherwise it is finite.

In the next section we will choose the symmetric gauge $\chi=$constant, that
is 
\end{mathletters}
\begin{equation}
(A_{1}, A_{2})=\frac{B}{2}(-q_{2}, q_{1}),
\end{equation}
and discuss the Wigner functions in any gauge in section XI. Before doing
that we should note that in a gauge which is at most quadratic function of $%
q_{j}$'s $a, \bar{a}, b, \bar{b}$ are some linear functions of $q_{j}$'s and 
$p_{j} $'s. Hence, by virtue of (2.10) we get 
\begin{mathletters}
\begin{eqnarray}
(\bar{a}_{\star})^{k}&=&\bar{a}^{k}, \quad ({a}_{\star})^{k}= a^{k},\\
(\bar{b}_{\star})^{k}&=&\bar{b}^{k},\quad ({b}_{\star})^{k}=b^{k},
\end{eqnarray}
\end{mathletters}
and rewrite (3.15) as
\begin{eqnarray}
W_{nl} = \frac{1}{n ! l !} \bar{a}^{n}\star \bar{b}^{l}\star W_{0} \star
a^{n}\star b^{l}.
\end{eqnarray}
Note that the $\star$-product between $a$'s and $b$'s can be removed for $%
\bar{a}\star \bar{b}=\bar{a}\bar{b}$ and $a\star b=ab$.

\section{The Ground State Wigner Function}

Let us introduce the complex coordinates 
\begin{mathletters}
\begin{eqnarray}
z &=&\frac{1}{\sqrt{2}}(q_{1}+iq_{2}),\quad p=\frac{1}{\sqrt{2}}%
(p_{1}-ip_{2}), \\
\bar{z} &=&\frac{1}{\sqrt{2}}(q_{1}-iq_{2}),\quad \bar{p}=\frac{1}{\sqrt{2}}%
(p_{1}+ip_{2}),
\end{eqnarray}
which imply 
\end{mathletters}
\begin{mathletters}
\begin{eqnarray}
\partial _{q_{1}} &=&\frac{1}{\sqrt{2}}(\partial _{z}+\partial _{\bar{z}%
}),\quad \partial _{p_{1}}=\frac{1}{\sqrt{2}}(\partial _{p}+\partial _{\bar{p%
}}), \\
\partial _{q_{2}} &=&\frac{i}{\sqrt{2}}(\partial _{z}-\partial _{\bar{z}%
}),\quad \partial _{p_{2}}=\frac{i}{\sqrt{2}}(\partial _{\bar{p}}-\partial
_{p}).
\end{eqnarray}
In terms of new coordinates the $\star $-product (1.3) (for $D=2$)
transforms to 
\end{mathletters}
\begin{equation}
\star =\exp [\frac{1}{2}i\hbar (\stackrel{\leftarrow }{\partial }_{z}%
\stackrel{\rightarrow }{\partial }_{p}+\stackrel{\leftarrow }{\partial }_{%
\bar{z}}\stackrel{\rightarrow }{\partial }_{\bar{p}}-\stackrel{\leftarrow }{%
\partial }_{p}\stackrel{\rightarrow }{\partial }_{z}-\stackrel{\leftarrow }{%
\partial }_{\bar{p}}\stackrel{\rightarrow }{\partial }_{\bar{z}})],
\end{equation}
and the annihilation functions take the form 
\begin{mathletters}
\begin{eqnarray}
a&=&\frac{\sqrt{2}}{m\gamma \omega }(\bar{p}-\frac{im\omega }{2}z),\\
b&=&\frac{\sqrt{2}}{m\gamma \omega }(-p+\frac{im\omega }{2}\bar{z}).
\end{eqnarray}
\end{mathletters}
In that case the defining relations (3.14) of $W_{0}$ are as follow 
\begin{mathletters}
\begin{eqnarray}
\bar{p}\star W_{0}-\frac{im\omega }{2}z\star W_{0} &=&0, \\
p\star W_{0}-\frac{im\omega }{2}\bar{z}\star W_{0} &=&0.
\end{eqnarray}
Combining the complex conjugate of (4.5a) with (4.5b) we get 
\end{mathletters}
\begin{equation}
\partial _{z}W_{0}=-\frac{m\omega }{\hbar }\bar{z}W_{0},\quad \partial _{%
\bar{p}}W_{0}=-\frac{4}{m\omega \hbar }pW_{0}.
\end{equation}
By ignoring a possible factor $f(\bar{z},p)$ which destroys the reality of $%
W_{0}$, and by defining 
\begin{equation}
H_{0}=\frac{1}{m}p\bar{p}+\frac{1}{4}m\omega ^{2}z\bar{z}=
\frac{{\bf p }^{2}}{2m}+\frac{m\omega ^{2}}{8}{\bf q}^{2},
\end{equation}
which corresponds the Hamiltonian function of a two-dimensional isotropic
harmonic oscillator, the general real solution of Eqs. (4.6) is 
\begin{equation}
W_{0}=4e^{-\frac{4}{\hbar \omega }H_{0}}.
\end{equation}

We have normalized $W_{0}$ as $\int_{{\bf R}^{4}}W_{0}dV=h^{2}$, where
the volume form $dV$ is
\begin{eqnarray}
dV =dq_{1}\wedge dq_{2}\wedge dp_{1}\wedge dp_{2} =dz\wedge d\bar{z}\wedge
dp\wedge d\bar{p}.
\end{eqnarray}
We note that the coordinate transformation given by (4.1) is a canonical
transformation since the symplectic 2-form $\Omega$ remains invariant: 
\begin{eqnarray}
\Omega = dq_{1}\wedge dp_{1}+ dq_{2}\wedge dp_{2} =dz\wedge dp+ d\bar{z}%
\wedge d\bar{p}.
\end{eqnarray}

\section{Wigner Functions of Higher States}

Observing that the $\star $-product (4.3) can be expressed in terms of $a, 
\bar{a}, b, \bar{b}$, explicit expressions of the Wigner functions and other
related computations can be carried out in an easier and more suggestive
way. To this end from Eqs. (4.4) and their complex conjugates we obtain 
\begin{mathletters}
\begin{eqnarray}
z &=&\frac{i\gamma }{\sqrt{2}}(a+\bar{b}),\quad p=\frac{m\omega \gamma }{%
2\sqrt{2}}(\bar{a}-b), \\
\bar{z} &=&-\frac{i\gamma }{\sqrt{2}}(\bar{a}+b),\quad \bar{p}=\frac{m\omega
\gamma }{2\sqrt{2}}(a-\bar{b}),
\end{eqnarray}
and 
\end{mathletters}
\begin{mathletters}
\begin{eqnarray}
\partial _{z} &=&-\frac{i}{\gamma \sqrt{2}}(\partial _{a}+\partial _{\bar{b}%
}),\quad \partial _{p}=\frac{\sqrt{2}}{m\omega \gamma }(\partial _{\bar{a}%
}-\partial _{b}), \\
\partial _{\bar{z}} &=&\frac{i}{\gamma \sqrt{2}}(\partial _{\bar{a}}+\partial
_{b}),\quad \partial _{\bar{p}}=\frac{\sqrt{2}}{m\omega \gamma }(\partial
_{a}-\partial _{\bar{b}}).
\end{eqnarray}
Then (4.3) and (4.7) can be written as 
\end{mathletters}
\begin{mathletters}
\begin{eqnarray}
\star &=&exp[\frac{1}{2}(\stackrel{\leftarrow }{\partial }_{a}\stackrel{%
\rightarrow }{\partial }_{\bar{a}}+\stackrel{\leftarrow }{\partial }_{b}%
\stackrel{\rightarrow }{\partial }_{\bar{b}}-\stackrel{\leftarrow }{\partial 
}_{\bar{a}}\stackrel{\rightarrow }{\partial }_{a}-\stackrel{\leftarrow }{%
\partial }_{\bar{b}}\stackrel{\rightarrow }{\partial }_{b}), \\
H_{0} &=&\frac{1}{2}\hbar \omega (\bar{a}a+\bar{b}b).
\end{eqnarray}

By inserting (5.3b) and (4.8) into (3.22) and then by defining 
\end{mathletters}
\begin{mathletters}
\begin{eqnarray}
w_{n}&=&\frac{1}{n!} \bar{a}^{n}\star e^{-2 a\bar{a} }\star a^{n},\\
w_{l}&=&\frac{1}{l!} \bar{b}^{l}\star e^{-2b\bar{b}}\star b^{l},
\end{eqnarray}
\end{mathletters}
$W_{nl}$ can be cast to the following factorized form
\begin{equation}
W_{nl}= 4w_{n}(a, \bar{a})w_{l}(b, \bar{b}).
\end{equation}
By virtue of (5.3a) one can easily verify that 
\begin{equation}
e^{-2a\bar{a}}\star a^{n}=\sum_{k=0}^{n}(-\frac{1}{2})^{k}\frac{1}{k!}%
(\partial _{\bar{a}}^{k}e^{-2a\bar{a}})(\partial
_{a}^{k}a^{n})=2^{n}a^{n}e^{-2a\bar{a}},
\end{equation}
where $(_{k}^{n})$ being a binomial number we have used the identity $%
\sum_{k=0}^{n}(_{k}^{n})=2^{n}$. On substituting (5.6) into (5.4a) we obtain 
\begin{eqnarray}
w_{n}&=&\frac{2^{n}}{n!}\bar{a}^{n}\star (a^{n}e^{-2a\bar{a}})\nonumber\\
&=&\frac{2^{n}}{n!}%
(\bar{a}-\frac{1}{2}\partial _{a})^{n}(a^{n}e^{-2a\bar{a}})=e^{-2a\bar{a}%
}P_{n}(2a\bar{a}),
\end{eqnarray}
where we have defined 
\begin{equation}
P_{n}(u)=\frac{1}{n!}e^{u}(1-\frac{d}{du})^{n}(u^{n}e^{-u}).
\end{equation}

In the appendix we will prove the following operator identity 
\begin{equation}
(1-\frac{d}{du})^{n}(u^{n}e^{-u})=(-1)^{n}e^{u}(\frac{d}{du}%
)^{n}(u^{n}e^{-2u}).
\end{equation}
On multiplying both sides of these identity by $e^{u}/n!$ and by recalling
the Rodriguez formula of the Laguerre polynomials (see the appendix) 
\begin{equation}
L_{n}(2u)=\frac{e^{2u}}{n!}\frac{d^{n}}{du^{n}}(u^{n}e^{-2u}),
\end{equation}
we see that 
\begin{equation}
P_{n}(u)=(-1)^{n}L_{n}(2u).
\end{equation}
Therefore, from (5.7) and (5.11) the explicit functional form of diagonal
Wigner functions (5.5) are obtained as follows 
\begin{equation}
W_{nl}=(-1)^{n+l}4L_{n}(4a\bar{a})L_{l}(4b\bar{b})e^{-2(a\bar{a}+b\bar{b})}.
\end{equation}

\section{Symmetry Properties}

Some transformation and symmetry properties of Wigner functions immediately
follow from their factorized forms given by (5.12) and from following
relations 
\begin{mathletters}
\begin{eqnarray}
\bar{a}a &=&\frac{H_{L}}{\hbar \omega }
= \frac{1}{2\hbar \kappa
}|p+i\kappa \bar{z}|^{2} \nonumber\\
&=&\frac{1}{4\hbar \kappa }%
[(p_{1}+\kappa q_{2})^{2}+(p_{2}-\kappa q_{1})^{2}],\\
\bar{b}b &=&\frac{J}{\hbar }+\frac{H_{L}}{\hbar \omega }
=\frac{1}{2\hbar
\kappa }|p-i\kappa \bar{z}|^{2}\nonumber\\
&=& \frac{1}{4\hbar
\kappa }[(p_{1}-\kappa q_{2})^{2}+(p_{2}+\kappa q_{1})^{2}],
\end{eqnarray}
where $\kappa =m\omega /2$. As is apparent from the first equalities of
(6.1a) and (6.1b), the values of $W_{nl}$ depend on the level sets of $%
H_{L}/\hbar \omega $ and of $J/\hbar $ which are classical constants of
motion and define some surfaces in ${\bf R^{4}}$. Rewriting
(5.12) as 
\end{mathletters}
\begin{equation}
W_{nl}=(-1)^{n+l}4L_{n}(4\frac{H_{L}}{\hbar \omega })L_{l}(4\frac{J}{\hbar }%
+4\frac{H_{L}}{\hbar \omega })e^{-2\frac{J}{\hbar }-4\frac{H_{L}}{\hbar
\omega }}.
\end{equation}
makes it possible to read out the values of $W_{nl}$'s on various level sets
very easily. In particular, since $L_{n}(0)=1$, at the phase-space origin
which corresponds to $H_{L}=0=J$ we have 
\begin{equation}
W_{nl}(0,0)=4(-1)^{n+l}.
\end{equation}

The following relations now easily follow from (5.12) and (6.1), 
\begin{mathletters}
\begin{eqnarray}
W_{nl}(-{\bf q},{\bf p})&=&W_{ln}({\bf q},{\bf p})\nonumber\\
&=&W_{nl}({\bf q},-{\bf p}),
\\
W_{nl}(-{\bf q},-{\bf p})&=&W_{nl}({\bf q},{\bf p}), \\
W_{nl}(q_{2},q_{1},p_{2},p_{1})&=&W_{ln}({\bf q},{\bf p}).
\end{eqnarray}
\end{mathletters}
(6.4a) shows that under the space inversion
$({\bf q}\rightarrow -{\bf q})$ and time reversal
$({\bf p}\rightarrow -{\bf p})$ transformations all Wigner
functions transform in the same way. In particular, Wigner
functions for which $n=l$ possess both space inversion and
time reversal symmetries separately and they are positive
valued. Eq. (6.4b) directly results from (6.4a) and it implies
that all Wigner functions are invariant under the phase-space
parity transformation
$({\bf q},{\bf p})\rightarrow (-{\bf q},-{\bf p})$. The transformation
property given by (6.4c) will be helpful in deriving marginal probability
densities (see Sec. X). We should note that although (6.1a) and
(6.1b) have some translational invariance properties they do not
lead to a translational symmetry of $W_{nl}$. Namely, the
following translational invariance of $H_{L}$:
\begin{eqnarray}
H_{L}({\bf q}+{\bf c},{\bf p}-\kappa{\bf c}^{\ast })=
H_{L}({\bf q},{\bf p}),\nonumber
\end{eqnarray}
where ${\bf c}=(c_{1},c_{2}),{\bf c}^{\ast }=(c_{2},-c_{1})$,  is
not compatible with  that of (6.1b).

Due to the fact that $W_{nl}$ depend on the products $a\bar{a},b\bar{b}$, a
larger class of symmetries can immediately be identified. Indeed, it is
evident that under transformations given by 
\begin{equation}
(a, b, \bar{a}, \bar{b})\rightarrow (ue^{i\xi }a, ve^{i\eta}b, u^{-1}
e^{-i\xi }\bar{a}, v^{-1}e^{-i\eta }\bar{b}),
\end{equation}
where $\xi ,\eta $ and $u\neq0, v\neq 0 $ are some real
parameters, all $W_{nl}$ remain invariant. Obviously, this
transformation also leaves $H_{L}$ and $J$ invariant. To express
this transformation in terms of $({\bf q}, {\bf p})$ coordinates, we
define two column matrices ${\bf x}$ and ${\bf y}$ by
${\bf x}^{T}=(a,b,\bar{a},\bar{b}), {\bf y}^{T}=(q_{1},q_{2},p_{1},p_{2})$%
, where ${\bf x}^{T}$ denotes the transpose of ${\bf x}$. We then
introduce two nonsingular matrices 
\begin{mathletters}
\begin{eqnarray}
A&=&\left( 
\begin{array}{cccc}
ue^{i\xi } & 0 & 0 & 0 \\ 
0 & ve^{i\eta } & 0 & 0 \\ 
0 & 0 & u^{-1}e^{-i\xi } & 0 \\ 
0 & 0 & 0 & v^{-1}e^{-i\eta }
\end{array}
\right),\\
B&=&\frac{\gamma }{2}\left(
\begin{array}{cccc}
i & -i & -i & i \\ 
1 & 1 & 1 & 1 \\ 
\kappa & -\kappa & \kappa & -\kappa \\ 
-i\kappa & -i\kappa & i\kappa & i\kappa
\end{array}
\right),
\end{eqnarray}
\end{mathletters}
having the determinants $detA=1, detB=-\gamma^{4}\kappa^{2}=-\hbar ^{2}$. In
this matrix notation the transformation (6.5) and relation between ${\bf y}$
and ${\bf x}$ simply read as 
\begin{equation}
{\bf x}^{\prime }=A{\bf x},\quad {\bf y}=B{\bf x}.
\end{equation}
In writing ${\bf y}=B{\bf x}$ we made use of Eqs. (4.1) and (5.1). Since $%
{\bf y}^{\prime}=B{\bf x}^{\prime}= BA{\bf x} $ we deduce, from (6.7) that
the transformation (6.5) can be written, in the canonical coordinates, as 
\begin{eqnarray}
{\bf y}^{\prime }&=&C{\bf y},\\
C&=&BAB^{-1}=\frac{1}{4}\left(
\begin{array}{cc}
M & -\kappa^{-1}N \\ 
\kappa N & M
\end{array}
\right), \nonumber
\end{eqnarray}
where $y^{\prime
T}=(q^{\prime}_{1},q^{\prime}_{2},p^{\prime}_{1},p^{\prime}_{2})$ and 
\begin{eqnarray}
M&=&\left( 
\begin{array}{cc}
c_{+} & -s_{-} \\ 
s_{-} & c_{+}
\end{array}
\right),\quad N=\left( 
\begin{array}{cc}
s_{+} & c_{-} \\ 
-c_{-} & s_{+}
\end{array}
\right) , \\
c_{\pm }&=&u_{+}\cos \xi \pm v_{+}\cos \eta +i( u_{-}\sin \xi \pm v_{-}\sin
\eta),  \nonumber \\
s_{\pm }&=&u_{+}\sin \xi \pm v_{+}\sin \eta -i( u_{-}\cos \xi \pm v_{-}\cos
\eta), \\
u_{\pm}&=&u\pm \frac{1}{u},\quad v_{\pm}=v\pm \frac{1}{v}.  \nonumber
\end{eqnarray}

In view of $det C=det A=1$ and 
\begin{mathletters}
\begin{eqnarray}
NN^{T}+MM^{T}&=&16 {\bf 1},\\
NM^{T}-MN^{T}&=&{\bf 0},
\end{eqnarray}
\end{mathletters}
where ${\bf 1}$ and ${\bf 0}$ denote, respectively, $2\times 2$ unit and
zero matrices, it is straightforward to check that $C$ is a symplectic
matrix. That is, 
\begin{eqnarray}
J_{0}=\left( 
\begin{array}{cc}
{\bf 0} & {\bf 1} \\ 
-{\bf 1} & {\bf 0}
\end{array}
\right),\nonumber
\end{eqnarray}
being the standard symplectic matrix, $C$ satisfies the symplectic condition 
$CJ_{0}C^{T}=J_{0}$. Hence $C$ belongs to the symplectic group $Sp(4, {\bf C}%
)$. Evidently, $W_{nl}(C_{1}{\bf y})=W_{nl}({\bf y}) = W_{nl}(C_{2}{\bf y})$
imply that $W_{nl}(C_{1}C_{2}^{-1}{\bf y})=W_{nl}({\bf y})$. In other words,
if $C_{1}$ and $C_{2}$ matrices leave $W_{nl}({\bf y})$ invariant so does $%
C_{1}C_{2}^{-1}$. This means that the transformations (6.8) form a four
parameter subgroup of $Sp(4, {\bf C})$.

Finally in this section we note that the volume form and symplectic
2-form in $(a,b,\bar{a},\bar{b})$ coordinates take the form 
\begin{mathletters}
\begin{eqnarray}
dV &=&dq_{1}\wedge dq_{2}\wedge dp_{1}\wedge dp_{2}=\hbar ^{2}da\wedge d\bar{%
a}\wedge db\wedge d\bar{b}, \\
\Omega &=&dq_{1}\wedge dp_{1}+dq_{2}\wedge dp_{2}=i\hbar (da\wedge d\bar{a}%
+db\wedge d\bar{b}).
\end{eqnarray}
Then, the symplectic property of the transformation (6.5) can directly be
observed from the second equality of (6.12b).

\section{Generic properties}

In this section we shall exhibit some generic properties of diagonal Wigner
functions, such as the reality, $\star $-projection, normalization, and
orthogonality properties, within autonomous framework of deformation
quantization. Marginal probability densities and their basic properties will
be taken up in section X. As is apparent from (2.3), (3.15) (or (3.22)) and
(4.8) the reality condition $\overline{W}_{nl}=W_{nl}$ of diagonal Wigner
functions is guaranteed by construction from the beginning. On the other
hand, the following form of the so-called $\star $-projection property 
\end{mathletters}
\begin{equation}
W_{nl}\star W_{n^{\prime }l^{\prime }}\propto \delta _{nn^{\prime }}\delta
_{ll^{\prime }}W_{nl}
\end{equation}
can be easily inferred from (3.18) by taking the $\star$-multiplication of
them from the left and then from the right by $W_{n^{\prime }l^{\prime }}$.
We start by deriving the exact form of this relation for the Wigner function
of Landau levels.

Noting that, by (5.3a) 
\begin{equation}
(a\bar{a})^{k+1} = (a\bar{a})^{k}\star a\bar{a}+\frac{k^{2}}{4}
(a\bar{a})^{k-1}
\end{equation}
one can verify $(a\bar{a})^{k}\star W_{0}=
(k!/2^{k})W_{0}$ by induction and
\begin{equation}
e^{-2ta\bar{a}}\star W_{0}=\frac{1}{1+t }W_{0}.
\end{equation}
This is a special case of $e^{-2ta\bar{a}}\star e^{-2sa\bar{a}}
=(1+ts)^{-1} \exp[-2(t+s)(1+ts)^{-1} a\bar{a}]$ which can be proved
by other means (see Eq. 34 in \cite{Curtright}). Hence, we obtain
the $\star$-projection property of $W_{0}$
as follows\footnote{In \cite{Vercin2} a misleading factor $4/e^{2}$
that appears at the end of Eq. (7.4) and in
Eqs. (7.8), (7.9), (7.11) should be removed.}
\begin{eqnarray}
W_{0}\star W_{0} = 4e^{-2a\bar{a}}\star e^{-2b\bar{b}}\star W_{0}
=W_{0}.
\end{eqnarray}

By virtue of (3.13) we have 
\begin{eqnarray}
a^{n}\star \bar{a}^{n^{\prime}}&=& a^{n-1}\star \bar{a}^{n^{\prime}-1}\star
(N_{a}+n^{\prime})\nonumber\\
&=& a^{n-2}\star \bar{a}^{n^{\prime}-2}\star
(N_{a}+n^{\prime}-1)\star (N_{a}+n^{\prime}).
\end{eqnarray}
This leads us to, by induction 
\begin{eqnarray}
a^{n}\star \bar{a}^{n^{\prime}}= \left\{ 
\begin{array}{cc}
\bar{a}^{n^{\prime}-n}\star \prod ^{n^{\prime}}_{j=n^{\prime}-n+1}
(N_{a}+j)_{\star}, & {\rm for}\quad n^{\prime} >n, \\ 
{a}^{n-n^{\prime}}\star \prod ^{n^{\prime}}_{j=1} (N_{a}+j)_{\star}, & 
{\rm for}\quad n^{\prime} < n, \\ 
\prod ^{n}_{j=1} (N_{a}+j)_{\star}, & {\rm for}\quad n^{\prime} = n,
\end{array}
\right.
\end{eqnarray}
where 
\begin{eqnarray}
\prod^{n^{\prime}}_{j=i} (N_{a}+j)_{\star}\equiv (N_{a}+i)\star (N_{a} +
i+1)\star \cdots \star (N_{a}+n^{\prime}).
\end{eqnarray}
Similar relations hold for $b^{l}\star \bar{b}^{l^{\prime}}$. On the other
hand since 
\begin{eqnarray}
W_{0}\star N_{a}\star W_{0} =0 = W_{0}\star N_{b}\star W_{0},  \nonumber \\
W_{0}\star a^{k}\star W_{0} =0 = W_{0}\star \bar{a}^{k}\star W_{0}, 
\nonumber \\
W_{0}\star b^{k}\star W_{0} =0 = W_{0}\star \bar{b}^{k}\star W_{0}, 
\nonumber
\end{eqnarray}
we see that 
\begin{eqnarray}
I_{nn^{\prime},ll^{\prime}}&\equiv & W_{0}\star a^{n}\star \bar{a}%
^{n^{\prime}}\star b^{l}\star \bar{b}^{l^{\prime}}\star W_{0} \nonumber \\
&=&n!l!
\delta_{nn^{\prime}}\delta_{ll^{\prime}} W_{0} \star W_{0}
= n!l!\delta_{nn^{\prime}}\delta_{ll^{\prime}} W_{0}.
\end{eqnarray}
Then, from (3.22) and (7.8) we obtain
\begin{equation}
W_{nl}\star W_{n^{\prime}l^{\prime}}= \frac{1}{n!l! n^{\prime}!l^{\prime}!} 
\bar{a}^{n}\star \bar{b}^{l}\star I_{nn^{\prime},ll^{\prime}} \star
a^{n^{\prime}}\star b^{l^{\prime}}
= \delta_{nn^{\prime}}%
\delta_{ll^{\prime}} W_{nl},
\end{equation}
for all diagonal Wigner functions of Landau levels 

By virtue of (2.2), (3.22), (4.9) and (7.6) we have the following
normalization property
\begin{eqnarray}
\int_{{\bf R}^{4}}W_{nl}dV &=&\frac{1}{n!l!}\int_{{\bf R}^{4}}\bar{a}%
^{n}\star \bar{b}^{l}\star W_{0}\star a^{n}\star b^{l}dV,  \nonumber \\
&=&\frac{1}{n!l!}\int_{{\bf R}^{4}}a^{n}\star \bar{a}^{n}\star b^{l}\star 
\bar{b}^{l}\star W_{0}dV,  \nonumber \\
&=&\int_{{\bf R}^{4}}W_{0}dV=h^{2}.
\end{eqnarray}
Finally, the $\star$-orthogonality directly follows from (2.2), (7.9) and
(7.10): 
\begin{eqnarray}
\int_{{\bf R}^{4}}W_{nl} \star W_{n^{\prime}l^{\prime}}dV &=& \int_{{\bf R}%
^{4}}W_{nl} W_{n^{\prime}l^{\prime}}dV,  \nonumber \\
&=& \delta_{nn^{\prime}}\delta_{ll^{\prime}} \int_{{\bf R}%
^{4}} W_{nl}dV=h^{2} \delta_{nn^{\prime}}\delta_{ll^{\prime}}.
\end{eqnarray}
As is apparent, many properties of $W_{nl}$ are guaranteed by the
corresponding properties of $W_{0}$.

\section{Generating Function for all Wigner Functions}

Let us define the phase-space functions 
\begin{mathletters}
\begin{eqnarray}
G_{1}(a, \bar{a}) &=& e^{\alpha_{1} \bar{a}}\star e^{-2%
\bar{a} a}\star e^{\beta_{1}a}, \\
G_{2}(b, \bar{b}) &=& e^{\alpha_{2} \bar{b}}\star e^{-2%
\bar{b} b}\star e^{\beta_{2}b},
\end{eqnarray}
where $\bbox{\alpha}=(\alpha_{1}, \alpha_{2}), \bbox{\beta}=(\beta_{1},
\beta_{2})$ are some parameters. Using the shorthands
$G_{1}=G_{1}(a, \bar{a})$ and
$G_{2}=G_{2}(b, \bar{b})$ we define and compute
$w_{n_{1}n_{2}}(a, \bar{a}), w_{l_{1}l_{2}}(b, \bar{b})$ as follows 
\end{mathletters}
\begin{mathletters}
\begin{eqnarray}
w_{n_{1}n_{2}}(a, \bar{a})&=&
\sqrt{1/n_{1}!n_{2}!}\partial^{n_{1}}_{\alpha_{1}}
\partial^{n_{2}}_{\beta_{1}}G_{1}|_{\alpha_{1}=0=\beta_{1}},  \nonumber \\
&=&\sqrt{1/n_{1}!n_{2}!} \bar{a}^{n_{1}}\star e^{-2a\bar{a}}\star a^{n_{2}},
\\
w_{l_{1}l_{2}}(b, \bar{b})&=&\sqrt{1/l_{1}!l_{2}!}
\partial^{l_{1}}_{\alpha_{2}} \partial^{l_{2}}_{\beta_{2}} G_{2}|_{\alpha_{2} =0=\beta_{2}},  \nonumber \\
&=&\sqrt{1/l_{1}!l_{2}!} \bar{b}^{l_{1}}\star e^{-2b\bar{b}}\star b^{l_{2}}.
\end{eqnarray}
We then define the generating function 
\end{mathletters}
\begin{equation}
G= G_{1}(a, \bar{a}) G_{2}(b,\bar{b}),
\end{equation}
from which all (diagonal and off-diagonal) Wigner functions can be
generated as
\begin{eqnarray}
W_{n_{1}n_{2}l_{1}l_{2}}&=&\frac{4}{\sqrt{ n_{1}!n_{2}!l_{1}!l_{2}}}
\partial_{\alpha_{1}}^{n_{1}} \partial_{\beta_{1}}^{n_{2}}
\partial_{\alpha_{2}}^{l_{1}} \partial_{\beta_{2}}^{l_{2}}
G |_{\bbox{ \alpha}=\bbox{0}=\bbox{\beta}},  \nonumber \\
&=&4w_{n_{1}n_{2}}(a, \bar{a}) w_{l_{1}l_{2}}(b, \bar{b}),
\end{eqnarray}
In the case of $n_{1}=n_{2}=n, l_{1}=l_{2}=l$, (8.2) and (8.4)
coincide with (5.4) and (5.5).

Making use of 
\begin{eqnarray}
e^{\alpha_{1} \bar{a}}\star e^{-2\bar{a}a}=
e^{\alpha_{1} \bar{a}}e^{-\frac{1}{2}%
\stackrel{\leftarrow }{\partial }_{\bar{a}}\stackrel{\rightarrow }{\partial }%
_{a}}e^{-2\bar{a}a}=e^{2\bar{a}(\alpha_{1}-a)}
\end{eqnarray}
one can easily verify that 
\begin{eqnarray}
G_{1}=e^{-\alpha _{1}\beta _{1}}e^{2(\alpha _{1}\bar{a}%
+\beta _{1}a)}e^{-2\bar{a}a},
\end{eqnarray}
and a similar relation for $G_{2}$. Hence 
\begin{equation}
G=e^{-\bbox{\alpha \cdot \beta}} e^{2(\alpha _{1}\bar{a%
}+\beta _{1}a+\alpha _{2}\bar{b}+ \beta _{2}b)}e^{-\frac{4H_{0}}{\hbar
\omega }},
\end{equation}
where $\bbox{\alpha\cdot\beta}=\alpha _{1}\beta_{1}+\alpha _{2}\beta _{2}$
and we have used (5.3b).

Now let us consider the following expansion 
\begin{equation}
e^{-\alpha_{1} \beta_{1}}e^{2(\alpha_{1}\bar{a}+\beta_{1} a)}
=\sum_{k=0}^{\infty} \frac{\alpha_{1}^{k}}{k!}(2\bar{a})^{k} (1-\frac{%
\beta_{1}}{2\bar{a}})^{k} e^{2\beta_{1} a}.
\end{equation}
Making use of the generating function of the generalized Laguerre
polynomials \cite{Magnus,Gradshteyn} 
\begin{equation}
(1+y)^{k}e^{-xy}=\sum_{n=0}^{\infty}L_{n}^{k-n}(x)y^{n},
\end{equation}
which holds for $|y|<1$, we rewrite (8.8) as 
\begin{equation}
e^{-\alpha_{1}\beta_{1}}e^{2(\alpha_{1}\bar{a}+\beta_{1} a)} =\sum_{k,
n=0}^{\infty} \frac{\alpha_{1}^{k}}{k!}(2\bar{a})^{k-n}
(-\beta_{1})^{n}L_{n}^{k-n}(4a\bar{a}).
\end{equation}
The condition $|y|<1$ impose the constraint $2|\bar{a}|>|\beta_{1}|$ for Eq.
(8.10). Therefore, from (8.6) we have 
\begin{mathletters}
\begin{eqnarray}
G_{1} &=& e^{-2a\bar{a}}\sum_{k,n=0}^{\infty }\frac{%
\alpha _{1}^{k}}{k!}(2\bar{a})^{k-n}(-\beta _{1})^{n}L_{n}^{k-n}(4a\bar{a}),
\\
G_{2} &=&e^{-2b\bar{b}}\sum_{k,n=0}^{\infty }\frac{%
\alpha _{2}^{k}}{k!}(2\bar{b})^{k-n}(-\beta _{2})^{n}L_{n}^{k-n}(4b\bar{b}),
\end{eqnarray}
and, by (8.2) 
\end{mathletters}
\begin{mathletters}
\begin{eqnarray}
w_{n_{1}n_{2}}(a,\bar{a}) &=&\sqrt{\frac{n_{2}!}{n_{1}!}}(-1)^{n_{2}}(2\bar{%
a})^{n_{1}-n_{2}}L_{n_{2}}^{n_{1}-n_{2}}(4a\bar{a})e^{-2a\bar{a}}, \\
w_{l_{1}l_{2}}(b,\bar{b}) &=&\sqrt{\frac{l_{2}!}{l_{1}!}}(-1)^{l_{2}}(2\bar{%
b})^{l_{1}-l_{2}}L_{l_{2}}^{l_{1}-l_{2}}(4b\bar{b})e^{-2b\bar{b}}.
\end{eqnarray}

On substituting (8.12) into (8.4) we get 
\end{mathletters}
\begin{equation}
W_{n_{1}n_{2}l_{1}l_{2}}= 4\sqrt{\frac{n_{2}!l_{2}!}{n_{1}!l_{1}!}}
(-1)^{n_{2}+l_{2}}(2\bar{a})^{n_{1}-n_{2}}(2\bar{b}%
)^{l_{1}-l_{2}}L_{n_{2}}^{n_{1}-n_{2}}(4a\bar{a})L_{l_{2}}^{l_{1}-l_{2}}(4b%
\bar{b})e^{-\frac{4H_{0}}{\hbar \omega }}.
\end{equation}
We should note that when superscript of a Laguerre polynomials is a negative
integer the following formula must be used 
\begin{equation}
L_{n}^{-k}(x)=(-x)^{k}\frac{(n-k)!}{n!}L_{n-k}^{k}(x),
\end{equation}
where $k$ is positive integer. In such a case the resulting expression can
also be obtained by starting with the following equivalent expansion of the
left hand side of (8.8): 
\[
e^{-\alpha _{1}\beta _{1}}e^{2(\alpha _{1}\bar{a}+\beta
_{1}a)}=\sum_{k=0}^{\infty }\frac{\beta _{1}^{k}}{k!}(2a)^{k}(1-\frac{\alpha
_{1}}{2a})^{k}e^{2\alpha _{1}\bar{a}}. 
\]
In any case (8.13) coincides with (5.12) for $n_{1}=n_{2}=n$ and $%
l_{1}=l_{2}=l$.

\section{ Phase-Space Coherent States and Generating Functions for Marginal
Probability Densities}

In this section we shall exhibit two important properties of the
generating function $G$. The first is that $G$ can be interpreted
as a phase-space coherent state of Landau levels. The second property
is that the integrated forms of $G$ over some phase-space planes
serve as generating functions for marginal probability densities (distributions)
on these planes.

For the first property we compute 
\begin{mathletters}
\begin{eqnarray}
a\star G &=&(a+\frac{1}{2}\partial _{\bar{a}})G
=\alpha _{1}G, \\
b\star G &=&(b+\frac{1}{2}\partial _{\bar{b}})G
=\alpha _{2}G,
\end{eqnarray}
which imply that $G$ is a left (complex) $\star$%
-eigenfunction of the annihilation functions. On the other hand, since 
\end{mathletters}
\begin{mathletters}
\begin{eqnarray}
G\star \bar{a} &=&(\bar{a}+\frac{1}{2}\partial _{a})G
=\beta _{1}G, \\
G\star \bar{b} &=&(\bar{b}+\frac{1}{2}\partial _{b})G
=\beta _{2}G,
\end{eqnarray}
$G$ is at the same time the right $\star$%
-eigenfunction of the creation functions. For these reasons $G$
behaves as a left/right coherent state. Moreover
\end{mathletters}
\begin{mathletters}
\begin{eqnarray}
\bar{a}\star G &=&(\bar{a}-\frac{1}{2}\partial _{a})G
=(2\bar{a}-\beta _{1})G=\partial
_{\alpha _{1}}G, \\
\bar{b}\star G &=&(\bar{b}-\frac{1}{2}\partial _{b})G
=(2\bar{b}-\beta _{2})G=\partial
_{\alpha _{2}}G, \\
G\star a &=&(a-\frac{1}{2}\partial _{\bar{a}})G
=(2a-\alpha _{1})G=\partial _{\beta
_{1}}G, \\
G\star b &=&(b-\frac{1}{2}\partial _{\bar{b}})G
=(2b-\alpha _{2})G=\partial _{\beta
_{2}}G.
\end{eqnarray}
provide us with the phase-space Bopp realization of left (right) $\star$%
-actions of $\bar{a}, \bar{b}$ ($a, b$). To make $G$
real it is sufficient to take $\bar{\alpha}=\beta $. In such a case
$G$ corresponds to the Glauber-Perelomov standard coherent
state\cite{Perelomov}.

To exhibit the second property of $G$, we define the
generating function for the marginal probability densities in the $%
q_{1}q_{2}$-plane as follows 
\end{mathletters}
\begin{equation}
M_{\alpha \beta}(q_{1},q_{2}) = \int_{-\infty}^{\infty}
\int_{-\infty}^{\infty}Gdp_{1}dp_{2}.
\end{equation}
Then the marginal probability density $P_{nl}(q_{1}, q_{2})$ in the $%
q_{1}q_{2}$-plane can be generated as 
\begin{equation}
P_{nl}(q_{1}, q_{2})=\frac{4}{n!l!} (\partial_{\alpha_{1}}
\partial_{\beta_{1}})^{n} (\partial_{\alpha_{2}}\partial_{\beta_{2}})^{l}
M_{\alpha \beta}(q_{1},q_{2}) |_{\bbox{\alpha}=\bbox{0}=\bbox{\beta}}.
\end{equation}
Combining (9.4) and (9.5) and comparing the result with (8.4) we obtain 
\begin{equation}
P_{nl}(q_{1}, q_{2})=\int_{-\infty}^{\infty}
\int_{-\infty}^{\infty}W_{nl}dp_{1}dp_{2},
\end{equation}
provided that we can change the order of  derivatives with respect
to parameters with integration on phase-space coordinates. The relation
(9.6) justifies our naming $M_{\alpha \beta}$ as the generating
functions for marginal probability densities (see also (11.2)).

Similarly, the generating function for the marginal probability
density in the $p_{1}p_{2}$-plane can be defined as 
\begin{equation}
M_{\alpha \beta}(p_{1},p_{2}) = \int_{-\infty}^{\infty}
\int_{-\infty}^{\infty}Gdq_{1}dq_{2},
\end{equation}
and the marginal probability density $P_{nl}(p_{1}, p_{2})$ in the $%
p_{1}p_{2}$-plane can be generated as 
\begin{equation}
P_{nl}(p_{1}, p_{2})=\frac{4}{n!l!} (\partial_{\alpha_{1}}
\partial_{\beta_{1}})^{n} (\partial_{\alpha_{2}}\partial_{\beta_{2}})^{l}
M_{\alpha \beta}(p_{1},p_{2}) |_{{\bf \alpha =0=\beta}}.
\end{equation}
Generalizing these observations we can define generating functions $%
M_{\alpha \beta}(q_{j},p_{k}); j, k=1, 2$ in order to generate the marginal
densities $P_{nl}(q_{j}, p_{k})$ on the $q_{j}p_{k}$-planes.

\section{Marginal Probability Densities of Landau Levels}

To ease the calculations of this section we shall use the dimensionless
coordinates 
\begin{mathletters}
\begin{eqnarray}
Z &=& \frac{1}{\gamma}(q_{1}+ iq_{2}),\quad \bar{Z} =\frac{1}{\gamma} (
q_{1}- iq_{2}), \\
P &=&\frac{1}{m\omega\gamma} (p_{1}+ ip_{2}),\quad \bar{P} = \frac{1}{%
m\omega\gamma}(p_{1}- ip_{2}),
\end{eqnarray}
and define the dimensionless quantities 
\end{mathletters}
\begin{mathletters}
\begin{eqnarray}
\rho^{2} &=& Z\bar{Z}=\frac{1}{\gamma^{2}}(q_{1}^{2}+q_{2}^{2}), \\
\zeta^{2} &=& P\bar{P}=
\frac{\gamma^{2}}{4\hbar^{2}}(p_{1}^{2}+p_{2}^{2}), \\
I_{q}&=&(\alpha_{1}+\beta_{2})\bar{Z} -(\alpha_{2}+\beta_{1})Z, \\
I_{p}&=&(\alpha_{2}-\beta_{1})P -(\alpha_{1}-\beta_{2})\bar{P}.
\end{eqnarray}
We then rewrite (8.7) as 
\end{mathletters}
\begin{equation}
G= e^{-\bbox{\alpha \cdot \beta}} e^{-\rho^{2}+iI_{q}}
e^{-4\zeta^{2}-2I_{p}}.
\end{equation}

Substituting (10.3) into (9.4) we obtain 
\begin{equation}
M_{\alpha \beta }(q_{1},q_{2})=e^{-\bbox{\alpha \cdot \beta }}e^{-\rho
^{2}+iI_{q}}K_{1}K_{2}=\pi (\frac{\hbar }{\gamma })^{2}e^{-\rho
^{2}}Q_{\alpha }Q_{\beta }, 
\end{equation}
where 
\begin{eqnarray}
K_{1} &=&\int_{-\infty }^{\infty }e^{-\frac{\gamma ^{2}}{\hbar ^{2}}%
[p_{1}^{2}-\frac{2\hbar ^{2}}{m\omega \gamma ^{3}}(\alpha _{1}-\alpha
_{2}+\beta _{1}-\beta _{2})p_{1}]}dp_{1}\nonumber\\
&=&\pi ^{1/2}\frac{\hbar }{\gamma }e^{%
\frac{1}{4}(\alpha _{1}-\alpha _{2}+\beta _{1}-\beta _{2})^{2}},\nonumber\\
K_{2} &=&\int_{-\infty }^{\infty }e^{-\frac{\gamma ^{2}}{\hbar ^{2}}%
[p_{2}^{2}+i\frac{2\hbar ^{2}}{m\omega \gamma ^{3}}(\alpha _{1}+\alpha
_{2}-\beta _{1}-\beta _{2})p_{2}]}dp_{2}\nonumber\\
&=&\pi ^{1/2}\frac{\hbar }{\gamma }e^{-%
\frac{1}{4}(\alpha _{1}+\alpha _{2}-\beta _{1}-\beta _{2})^{2}},\nonumber
\end{eqnarray}
are obtained by the standard complex integration methods and 
\begin{equation}
Q_{\alpha }=e^{-\alpha _{1}\alpha _{2}+i(\alpha _{1}\bar{Z}-\alpha
_{2}Z)},\quad Q_{\beta }=e^{-\beta _{1}\beta _{2}-i(\beta _{1}Z-\beta _{2}%
\bar{Z})}. 
\end{equation}
When Eq. (10.4) is substituted into (9.5), in terms of 
\begin{equation}
J_{nl}^{1}=\partial _{\alpha _{1}}^{n}\partial _{\alpha _{2}}^{l}Q_{\alpha
}|_{\alpha _{1}=0=\alpha _{2}},\quad J_{nl}^{2}=\partial _{\beta
_{1}}^{n}\partial _{\beta _{2}}^{l}Q_{\beta }|_{\beta _{1}=0=\beta _{2}},
\end{equation}
we obtain 
\begin{equation}
P_{nl}(q_{1},q_{2})=\frac{4\pi }{n!l!}(\frac{\hbar }{\gamma })^{2}e^{-\rho
^{2}}J_{nl}^{1}J_{nl}^{2}.
\end{equation}

Similar to (8.8) and (8.10), we can expand (10.5) as follows 
\begin{mathletters}
\begin{eqnarray}
Q_{\alpha}&=& \sum_{j=0}^{\infty} \frac{(i\alpha_{1}\bar{Z})^{j}}{j!}
(1+i\frac{\alpha_{2}}{\bar{Z}})^{j} e^{-i\alpha_{2}Z}\nonumber\\
&=& \sum_{j,
k=0}^{\infty}\frac{i^{j+k}}{j!}\bar{Z}^{j-k}
\alpha_{1}^{j}\alpha_{2}^{k}L^{j-k}_{k}(\rho^{2}), \\
Q_{\beta} &=& \sum_{j=0}^{\infty} \frac{(-i\beta_{1}Z)^{j}}{j!}
(1-i\frac{\beta_{1}}{Z})^{j} e^{i\beta_{2}\bar{Z}}\nonumber\\
&=& \sum_{j, k=0}^{\infty}\frac{%
(-i)^{j+k}}{j!} Z^{j-k} \beta_{1}^{j}\beta_{2}^{k} L^{j-k}_{k}(\rho^{2}).
\end{eqnarray}
Then, by using these in (10.6) we have 
\end{mathletters}
\begin{equation}
J_{nl}^{1}= i^{n+l}l!\bar{Z}^{n-l}L^{n-l}_{l}(\rho^{2}),\quad J_{nl}^{2}=
(-i)^{n+l}l! Z^{n-l} L^{n-l}_{l}(\rho^{2}),
\end{equation}
and, by (10.7) 
\begin{equation}
P_{nl}(q_{1}, q_{2})= 4\pi\frac{l!}{n!} (\frac{\hbar}{\gamma}%
)^{2}\rho^{2(n-l)}e^{-\rho^{2}} [L^{n-l}_{l}(\rho^{2})]^{2}.
\end{equation}

The generating functions $M_{\alpha \beta }(p_{1},p_{2}), M_{\alpha \beta
}(q_{1}, p_{1}), M_{\alpha \beta }(q_{2}, p_{2})$ and the corresponding
marginal probability densities can be calculated in a similar way. These are
all calculated and presented together in the Tables I and II. But the
calculations for the $q_{1}p_{2}$ and $q_{2}p_{1}$-planes are to be
presented as they are a bit different.

From (10.3) we have 
\begin{eqnarray}
M_{\alpha \beta }(q_{1}, p_{2})&=&\int_{-\infty }^{\infty }\int_{-\infty
}^{\infty }Gdq_{2}dp_{1},  \nonumber \\
&=&\pi \hbar e^{\frac{1}{2}(\tau^{2}_{+}+\tau^{2}_{-})}e^{\frac{1}{2}[(\alpha _{1}+i\tau
_{-})^{2}+(\beta _{1}-i\tau _{-})^{2}+(\alpha _{2}-i\tau _{+})^{2}+(\beta
_{2}+i\tau _{+})^{2}]},
\end{eqnarray}
where $\tau _{\pm }=(m\omega q_{1}\pm 2p_{2})/m\omega \gamma $. We now
consider the identity 
\begin{eqnarray}
\partial _{\alpha _{1}}^{n} e^{\frac{1}{2}(\alpha _{1}+i\tau
_{-})^{2}}|_{\alpha _{1}=0} &=&(-i)^{n}\partial _{\tau_{-}}^{n}e^{\frac{1}{2}%
(\alpha _{1}+i\tau _{-})^{2}}|_{\alpha _{1}=0},  \nonumber \\
&=&(-i)^{n}\partial _{\tau_{-}}^{n}e^{-\frac{1}{2}\tau_{-}^{2}} =(\frac{i}{%
\sqrt{2}})^{n}e^{-\frac{1}{2}\tau_{-}^{2}}H_{n}(\frac{\tau _{-}}{\sqrt{2}}).
\end{eqnarray}
In passing to the last equality we have used the definition of the Hermite
polynomials 
\begin{equation}
H_{n}(x)=(-1)^{n}e^{x^{2}}\partial _{x}^{n}e^{-x^{2}}.
\end{equation}
Then by using (10.12), (10.13) and the other three relations obtained for
derivatives with respect to $\alpha _{2},\beta _{1}$ and $\beta _{2}$ in 
\begin{equation}
P_{nl}(q_{1},p_{2})=\frac{4}{n!l!}(\partial _{\alpha _{1}}\partial _{\beta
_{1}})^{n}(\partial _{\alpha _{2}}\partial _{\beta _{2}})^{l}M_{\alpha \beta
}(q_{1},p_{2})|_{\bbox{\alpha}=\bbox{0}=\bbox{\beta}},
\end{equation}
we arrive at 
\begin{equation}
P_{nl}(q_{1},p_{2})=\frac{4\pi \hbar}{n!l!2^{n+l}}
e^{-\frac{1}{2}(\tau^{2}_{+}+\tau^{2}_{-}) } [H_{n}(\frac{%
\tau _{-}}{\sqrt{2}})H_{l}(\frac{\tau _{+}}{\sqrt{2}})]^{2}.
\end{equation}

As a common property, all marginal probability densities of Landau levels
are positive on the corresponding phase-space plane. They determine the
wavefunctions on these planes, up to some unitary phase factors. Axial
symmetry on the respective planes of the first four marginal probability
densities are obvious from their explicit expressions given in Table II.
Also in functional dependence $P_{nl}(q_{1},p_{2})$ and $P_{nl}(q_{2},p_{1})$
differ from others. Marginal probability densities along
$q_{j}$, or $p_{j}$, or along some oblique phase-space
directions can be computed as well. Finally we note that $P_{nl}(q_{2},p_{1})
$ presented in the last row of the Table II easily follows from (10.15) by
virtue of (6.4c).

\section{Remarks}

In this section some remarks that remain outside of our main goals are
collected together.

\subsection{Comparison with Definitions}

For justification we would like to compare (4.8) and (10.10) with
expressions obtainable from the definition of the Wigner functions. In the
symmetric gauge (3.20) and in terms of plane polar coordinates $(r,\theta )$
the normalized wavefunctions and the corresponding energy eigenvalues of the
Landau levels are given by 
\begin{eqnarray}
\psi _{n_{r}j}(r,\theta ) &=&\frac{1}{\gamma }
\sqrt{\frac{n_{r}!}{\pi (n_{r}+|j|)!}}
\rho ^{|j|}e^{ij\theta }e^{-\rho ^{2}/2}L_{n_{r}}^{|j|}(\rho ^{2}),
\nonumber \\
E_{n_{r}j} &=&\hbar \omega (n_{r}+\frac{1}{2}+\frac{|j|-j)}{2}), \\
n_{r} &=&0,1,2,...,\quad j=0,\pm 1,\pm 2,...  \nonumber
\end{eqnarray}
where $r=\gamma \rho =(q_{1}^{2}+q_{2}^{2})^{1/2}$ and $n_{r},j$ stand for
the radial and angular momentum quantum numbers. Since in (10.10) it is
supposed that $n\geq l$, we can make the identifications $|j|=n-l$ and $%
n_{r}+|j|=n$, which imply $n_{r}=l$. We then have, from (10.10) and (11.1) 
\begin{equation}
P_{nl}(q_{1},q_{2})=h^{2}|\psi _{n_{r}j}(r,\theta )|^{2}.
\end{equation}
This expected result confirms Eq. (10.10) and our definition of marginal
probability densities in section IX. On the other hand, by using the
normalized ground state wavefunction $\psi _{0}({\bf q})=\psi _{00}(r,\theta
)=\gamma \pi ^{-1/2}\exp (-\rho ^{2}/2)$ in the definition 
\begin{equation}
W_{0}=\int_{{\bf R}^{2}}\psi _{0}({\bf q}+\frac{1}{2}{\bf y})\bar{\psi}_{0}(%
{\bf q}-\frac{1}{2}{\bf y})e^{-i\frac{{\bf y}\cdot {\bf p}}{\hbar }%
}dy_{1}dy_{2},
\end{equation}
we get 
\begin{equation}
W_{0}=\frac{e^{-\frac{4H_{0}}{\hbar \omega }}}{\gamma ^{2}\pi }
\int_{-\infty }^{\infty }e^{-\frac{1}{4\gamma ^{2}}(y_{1}+i\frac{%
2\gamma ^{2}}{\hbar }p_{1})^{2}}dy_{1}\int_{-\infty }^{\infty }e^{-\frac{1}{%
4\gamma ^{2}}(y_{2}+i\frac{2\gamma ^{2}}{\hbar }p_{2})^{2}}dy_{2}.
\end{equation}
Noting that, each integral in (11.4) is equal to $2\gamma \pi ^{1/2}$ we
obtain $W_{0}=4\exp (-4H_{0}/\hbar \omega )$ which is the same as (4.8).
However, as we have mentioned in the introduction, in attempt to find other
confirmations the resulting integrals do not seem to be so easy to cope with.

\subsection{Unitary Transformations in a Phase-Space}

Let us consider a unitary similarity transformation generated by the
phase-space function $U\equiv U({\bf q}, {\bf p})$ and its $\star$-inverse $%
U^{-1}=\overline{U}$: 
\begin{equation}
U\star U^{-1}=U^{-1}\star U=1.
\end{equation}
When such a transformation applied to a $\star$-eigenvalue equation $H\star
W_{\lambda} = W_{\lambda} \star H=E_{\lambda} W_{\lambda} $, the transformed
functions 
\begin{equation}
H^{\prime}\equiv U\star H \star U^{-1},\quad W^{\prime}_{\lambda} \equiv
U\star W_{\lambda} \star U^{-1},
\end{equation}
obey the same $\star$-eigenvalue equation 
\begin{equation}
H^{\prime}\star W^{\prime}_{\lambda} = W^{\prime}_{\lambda} \star H^{\prime}
= E_{\lambda} W^{\prime}_{\lambda} .
\end{equation}
In view of (2.3), (7.9) and (11.5) it is easy to verify that 
\begin{mathletters}
\begin{eqnarray}
\overline{W}^{\prime}_{\lambda} &=& U\star \overline{W}_{\lambda}\star U^{-1}, \\
W^{\prime}_{\lambda}\star W^{\prime}_{\lambda^{\prime}} -k W
^{\prime}_{\lambda} \delta_{\lambda \lambda^{\prime}} &=& U\star (
W_{\lambda} \star W_{\lambda^{\prime}} -k W_{\lambda} \delta_{\lambda
\lambda^{\prime}}) \star U^{-1},
\end{eqnarray}
where $k$ is a constant. That is, the reality and $\star$-projection
properties of $W_{\lambda} $ are preserved under the unitary similarity
transformations. Moreover, as is evident from 
\end{mathletters}
\begin{eqnarray}
\int_{{\bf R}^{2D}}W^{\prime}_{\lambda} dV &=& \int_{{\bf R}^{2D}} U\star
W_{\lambda}\star U^{-1}dV \nonumber \\
&=&\int_{{\bf R}^{2D}}W_{\lambda} \star (U^{-1}\star U) dV = \int_{{\bf R}%
^{2D}} W_{\lambda} dV,
\end{eqnarray}
normalization properties of $W_{\lambda} $ is also preserved. In (11.9) we
have used (2.2) and (11.5).

\subsection{Gauge Transformations in a Phase-Space}

Now let us consider 
\begin{eqnarray}
U_{q}=e^{i\frac{q}{c\hbar}\chi},\quad U^{-1}_{q} =\overline{U}_{q} = e^{-i\frac{q%
}{c\hbar}\chi},
\end{eqnarray}
where $U_{q} \equiv U({\bf q})$ and $\chi\equiv \chi({\bf q})$ is a real
valued function of the coordinates. Obviously $U_{q} \star U^{-1}_{q} =U_{q}
U^{-1}_{q} =1$. By using (1.3) with $D=2$ we also have 
\begin{equation}
U_{q} \star {\bf p} \star U^{-1}_{q} = {\bf p}- \frac{q}{c}{\bf \nabla}%
_{q}\chi,
\end{equation}
which implies, for the Landau Hamiltonian (3.1) 
\begin{eqnarray}
H^{\prime}_{L}&=&\frac{1}{2m}[ U_{q} \star ({\bf p}-\frac{q}{c}{\bf A}) \star U^{-1}_{q}
]\cdot [U_{q} \star ({\bf p}-\frac{q}{c}{\bf A}) \star U^{-1}_{q} ]\nonumber\\
&=&\frac{1}{2m}[{\bf p}-\frac{q}{c}({\bf A}+{\bf \nabla}\chi)]^{2},
\end{eqnarray}
where $H^{\prime}_{L}=U_{q} \star H_{L}\star U^{-1}_{q} $ and ${\bf \nabla}%
_{q}$ represents two dimensional gradient operator in $q_{1}, q_{2}$
variables. Eqs. (11.11) and (11.12) show that the unitary transformation by $%
U_{q} $ amounts to the so called gauge transformations of the second kind in 
${\bf p} $ and $H_{L} $. Hence, without changing their generic properties
the Wigner functions of Landau levels in any
gauge $\chi $ can be computed directly from 
\begin{equation}
W^{\prime}_{n_{1}n_{2} l_{1} l_{2} } = U_{q} \star W_{n_{1} n_{2} l_{1}
l_{2} }\star U^{-1}_{q},
\end{equation}
where $W_{n_{1} n_{2} l_{1} l_{2} }$ are given by (8.13). In that case the
transformation in (11.13) may be considered as a first kind gauge
transformation. We should emphasize that the two forms of $\star$-product
given by (1.3) and (5.3a) can equally well be used in (11.13) provided that
the arguments of functions are properly written.

\subsection{Implications for Wigner Functions}

By using (1.3) we compute the following useful relation in an arbitrary
dimension 
\begin{eqnarray}
f_{1}({\bf q})\star e^{-\frac{i}{\hbar}{\bf y}\cdot {\bf p}} \star f_{2}(%
{\bf q}) &=& [e^{-\frac{i}{\hbar } {\bf y}\cdot ({\bf p}+\frac{i\hbar}{2}%
{\bf \nabla}_{q})}f_{1}({\bf q})] \star f_{2}({\bf q}),  \nonumber \\
&=& f_{1}({\bf q}+\frac{1}{2}{\bf y})[ e^{-\frac{i}{\hbar}{\bf y}\cdot {\bf p%
}}\star f_{2}({\bf q})],  \nonumber \\
&=& f_{1}({\bf q}+\frac{1}{2}{\bf y}) f_{2}({\bf q}-\frac{1}{2}{\bf y}) e^{-%
\frac{i}{\hbar}{\bf y}\cdot {\bf p}},
\end{eqnarray}
where ${\bf \nabla}_{q}$ denotes $D$-dimensional gradient operator in $q_{i}$
variables. Taking $f_{1}=U_{q} $ and $f_{2}=U_{q}^{-1} $ in (11.14) yields 
\begin{equation}
U_{q}\star e^{-\frac{i}{\hbar}{\bf y}\cdot {\bf p}} \star U_{q}^{-1} = e^{i%
\frac{q}{c\hbar}\chi ({\bf q}+\frac{1}{2}{\bf y})} e^{-i\frac{q}{c\hbar}\chi
({\bf q}-\frac{1}{2}{\bf y})} e^{-\frac{i}{\hbar}{\bf y}\cdot {\bf p}}
\end{equation}
This implies, for $W_{\lambda_{1} \lambda_{2}}^{\prime}=U_{q}\star
W_{\lambda_{1} \lambda_{2}}\star U_{q} ^{-1} $, that 
\begin{eqnarray}
W_{\lambda_{1} \lambda_{2}}^{\prime}&=& \int_{{\bf R}^{D}}
\psi_{\lambda_{1}} ({\bf q}+\frac{1}{2}{\bf y}) \overline{\psi}%
_{\lambda_{2}}({\bf q}-\frac{1}{2}{\bf y}) U_{q} \star e^{-\frac{i}{\hbar}%
{\bf y}\cdot {\bf p}} \star U_{q} ^{-1} dV(y),  \nonumber \\
&=& \int_{{\bf R}^{D}} \psi_{\lambda_{1}}^{\prime} ({\bf q}+\frac{1}{2}{\bf y%
}) \overline{\psi}_{\lambda_{2}}^{\prime} ({\bf q}-\frac{1}{2}{\bf y}) e^{-%
\frac{i}{\hbar}{\bf y}\cdot{\bf p}}dV(y),
\end{eqnarray}
where $W_{\lambda_{1}\lambda_{2}}$ is a off-diagonal Wigner function defined
by (1.1) and $\psi_{\lambda_{j}}^{\prime}({\bf q})=e^{\frac{iq}{c\hbar}\chi(%
{\bf q})}\psi_{\lambda_{j}}({\bf q})=U_{q}\psi_{\lambda_{j}}({\bf q})$. If
we change our notation as $W_{\psi_{\lambda_{1}}\psi_{\lambda_{2}}}=W_{%
\lambda_{1} \lambda_{2}}$, Eq. (11.16) implies the following important
relation for gauge transformation of the Wigner function 
\begin{equation}
W_{\psi_{1} \psi_{2}}^{\prime}= U\star W_{\psi_{1}\psi_{2}} \star U^{-1}=
W_{\psi_{1}^{\prime} \psi_{2}^{\prime}} =W_{U_{q}\psi_{1},U_{q}\psi_{2}}.
\end{equation}
Finally we should note that, for $f_{1}=\psi_{1}$ and $f_{2}=\overline{\psi}%
_{2}$ Eq. (11.14) suggests the following redefinition of Wigner function 
\begin{equation}
W_{\psi _{1},\psi _{2}}=\int_{{\bf R}^{D}}\psi _{1}({\bf q})\star e^{-\frac{i%
}{\hbar }{\bf y}\cdot {\bf p}}\star \overline{\psi }_{2}({\bf q})dV(y),
\end{equation}
which enables us to interpret the map $W$ as a ``sesquilinear $\star $%
-Fourier transformation''.

\section{Conclusion}

In this paper we have presented a problem that can, with all of its
features, thoroughly be considered in the framework of deformation
quantization without making use of other quantization schemes. As a summary,
we have obtained Wigner functions of Landau levels by solving a two-sided $%
\star$-eigenvalue equation, specified a large class of its transformation
and symmetry properties as well as established its generic properties in the
same framework. Off-diagonal Wigner functions and marginal probability
densities are generated by means of generating functions.

With its distinguishable and intriguing properties, Wigner function
provides the most complete description and offers convenient
interpretation of the result of several recent experiments in
atomic optics, molecular physics and signal processing
\cite{Nature,Leibfried,Kim}. A basic
aim of all these experiments is to probe the fundamental structure and
predictions of quantum mechanics in a new way, by observing non-classical
behaviors of Wigner function \cite{Bracken}. For this purpose marginal
probability densities play equally important role since Wigner function are
reconstructed from measured densities along various phase-space directions .
Now explicit expressions of Wigner functions and all marginal probability
densities of Landau levels at hand we do expect that the results of this
paper will shed more light on future experiments in which Landau levels are
realized and on their interpretations.

\acknowledgments

We thank M. \"{O}nder for his interest in this study and for many fruitful
discussions. This work was supported in part by the Scientific and Technical
Research Council of Turkey (T\"{U}B\.{I}TAK).

\appendix

\section{The Operator Identity (5.9)}

In this appendix we will prove the following operator identity (see
Eq.(5.9)) by induction: 
\begin{equation}
(1-T)^{n}(x^{n}e^{-x})=(-1)^{n}e^{x}T^{n}(x^{n}e^{-2x}),
\end{equation}
where $T=d/dx$ and $n$ is a positive integer. Then, we will present a finite
sum formula for the generalized Laguerre polynomials that follows from (A1).

Eq. (A1) holds for $n=1$ (and obviously for $n=0$). Now we suppose (A1) for $%
n-1$ and try to prove it for $n$. Defining $I_{k}=(1-T)^{k}(x^{k}e^{-x})$,
the inductive hypothesis is 
\begin{equation}
I_{n-1}= (-1)^{n-1} e^{x}T^{n-1} (x^{n-1}e^{-2x}).
\end{equation}
Then we have 
\begin{eqnarray}
I_{n} &=&(1-T)^{n}[x(x^{n-1}e^{-x})]= [x(1-T)-n]I_{n-1},  \nonumber \\
&=& (-1)^{n}e^{x}[xT^{n}+nT^{n-1}](x^{n-1}e^{-2x}), \\
&=& (-1)^{n}e^{x}T^{n}(x^{n}e^{-2x}).  \nonumber
\end{eqnarray}
In passing to the second equality we made use of $[(1-T)^{n}, x] =
-n(1-T)^{n-1}$, where $[,]$ stands for the usual commutator. The second line
of (A3) easily follows from (A2) and in passing to the last line we have
used 
\begin{equation}
(xT^{n}+nT^{n-1})f=T^{n}(xf),
\end{equation}
where $f$ is an arbitrary differentiable function of $x$. Eq. (A4) easily
results from the commutation relation $\lbrack T^{k}, x] = kT^{k-1}$, or
equivalently from the Leibniz rule $T^{n}(xf)=%
\sum_{k=0}^{n}(^{n}_{k})(T^{k}x)T^{n-k}f$. We should note that (A1) can be
generalized as 
\begin{equation}
(1-T)^{n}(fe^{x})=(-1)^{n}e^{x}T^{n}f,
\end{equation}
which can be proved more easily, again, by induction on $n$. (A1)
corresponds to choice $f=x^{n}e^{-2x}$ in (A5).

It is worth mentioning that, in view of the binomial expansion 
\begin{equation}
(1-T)^{n}=\sum_{j=0}^{n}(_{j}^{n})(-1)^{n-j} T^{n-j},
\end{equation}
and by recalling the Rodrigues formula for generalized Laguerre polynomials
( \cite{Magnus}, p. 241) 
\begin{equation}
L^{\alpha}_{n}(x)=\frac{e^{x}}{n!}x^{-\alpha}T^{n}(x^{n+\alpha}e^{-x}),
\end{equation}
from (A1) and (A7) we obtain the following finite sum formula 
\begin{equation}
L_{n}(2x)=\sum_{j=0}^{n} \frac{(-x)^{j}}{j!} L^{j}_{n-j}(x)
=\sum_{j=0}^{n}L^{-j}_{j}(x) L^{j}_{n-j}(x),
\end{equation}
where $L^{-j}_{j}(x)=(-x)^{j}/j!$. Since $L_{n}=L^{0}_{n} $, from (A7) we
get (5.10) for $\alpha=0$ and $x=2u$. We could not find (A8) in the related
classical references \cite{Magnus,Gradshteyn}.

\end{document}